\documentclass[10pt]{iopart}
\usepackage{latexsym,hyperref,color,amssymb,graphicx}

\newcommand{\beq}{\begin{equation}}
\newcommand{\eeq}{\end{equation}}
\newcommand{\bea}{\begin{eqnarray}}\newcommand{\eea}{\end{eqnarray}}
\newcommand{\bean}{\begin{eqnarray*}}
\newcommand{\eean}{\end{eqnarray*}}
\newcommand{\bei}{\begin{itemize}}
\newcommand{\eei}{\end{itemize}}
\newcommand{\ben}{\begin{enumeration}}
\newcommand{\een}{\end{enumeration}}
\newcommand{\nn}{\nonumber}

\definecolor{darkorange}{rgb}{.6,.2,.0}

\definecolor{darkgreen}{rgb}{0.0,0.7,0.0}

\begin{document}
\title{Rotations and Statistics}
\author{Kevin Cahill}
\address{New Mexico Center for Particle Physics, 
Department of Physics and Astronomy, 
University of New Mexico, Albuquerque, NM 87131}
\ead{cahill@unm.edu}
\date{\today}
\begin{abstract}
The way a field transforms under rotations
determines its statistics---as is easy to see 
for scalar, Dirac, and vector fields.
\end{abstract}
\maketitle

\section{Introduction\label{intro}}
At a Solvay conference
more than 40 years ago,
Eugene Wigner~\cite{Wigner1962} said that one
should be able to derive the connection between
spin and statistics~\cite{Fierz1939, Pauli1940}
directly from the properties of fields under rotations.
He noted that 
a suitable rotation \( U  \) by \(\pi\) about the origin
transforms
the mean value in
the vacuum of the product of
two scalar fields \( \phi(\mathbf{x},t) \phi(-\mathbf{x},t) \)
at the same time into that
of the fields in the opposite order 
\bea
\langle 0 | \phi(\mathbf{x},t) \phi(-\mathbf{x},t) | 0 \rangle
& = & \langle 0 | U^{-1} U \phi(\mathbf{x},t)
 U^{-1} U \phi(-\mathbf{x},t) U^{-1} U  | 0 \rangle \nn\\
& = & \langle 0 | \phi(- \mathbf{x},t) \phi(\mathbf{x},t) | 0 \rangle 
\label {Wigner}
\eea
so that the vacuum value of the commutator vanishes
\( \langle 0 | [ \phi(\mathbf{x},t), \phi(-\mathbf{x},t) ] | 0 \rangle = 0 \)\@.
What follows is an attempt to follow his suggestion;
I hope it will be of use to students.
\par
My basic assumptions are listed
in Section~\ref{assumptions}:
that quantum fields 
transform suitably under Lorentz transformations and 
are linear combinations 
of annihilation and creation operators
that satisfy either
commutation or anti-commutation
relations~\cite{WeinbergI4}\@.
In Sections~\ref{spin 0}--\ref{spin 1}, 
I show that scalar, Dirac, and vector fields
respectively commute, anti-commute, and commute at equal times
because of how they transform under rotations.
That their statistics extend to space-like separations
then follows from their properties under
Poincar\'{e} transformations.
In Sec.~\ref{any spin}, I show that a field \( \psi_{ab}(\mathbf{x},t) \)
that transforms under the \( (A,B) \) representation
of the Lorentz group will commute or anti-commute
with the field \( \psi_{ab}(\mathbf{y},t) \) of the same index
at equal times according to whether \( 2(A + B) \) 
is an even or an odd integer.

\section{Basic Assumptions\label{assumptions}}

\subsection{About Particles\label{About Particles}}
Since the particles of
a single species
are identical, the normalized states
\( | \dots; \mathbf{ p }, s; \mathbf{ p }', s'; \dots \rangle \)
and \( | \dots; \mathbf{ p }', s'; \mathbf{ p }, s; \dots \rangle \)
describe the same particles with the same sets of momenta
\( \dots, \mathbf{ p }, \mathbf{ p }', \dots \) 
and spin indices \( \dots, s , s', \dots \)\@.
Thus, these states 
differ at most by a phase factor
\beq
| \dots; \mathbf{ p }, s; \mathbf{ p }', s'; \dots \rangle =
e^{i \alpha} \, | \dots; \mathbf{ p }', s'; \mathbf{ p }, s; \dots \rangle.
\label {phase}
\eeq
In a space of three 
(but not two~\cite{PhysRevLett.48.1144,PhysRevLett.49.957})
spatial dimensions,
one may show~\cite{WeinbergI4} 
that the square of this phase factor is unity
so that
\beq
e^{i \alpha} = \pm 1 .
\label {pm 1}
\eeq
Thus~\cite{WeinbergI4} the (suitably normalized) 
annihilation and creation operators that
relate these states to each other and to the
vacuum satisfy either
commutation 
\beq
[ a( \mathbf{ p }, s), a( \mathbf{ p }', s') ] =  0 
\quad \mbox{and} \quad
[ a( \mathbf{ p }, s), a^\dagger( \mathbf{ p }', s') ] = 
\delta( \mathbf{ p } - \mathbf{ p }') \delta_{s s'} .
\label {[a,a]}
\eeq
or anti-commutation 
\beq
\{ a( \mathbf{ p }, s), a( \mathbf{ p }', s') \} =  0 
\quad \mbox{and} \quad
\{ a( \mathbf{ p }, s), a^\dagger( \mathbf{ p }', s') \} = 
\delta( \mathbf{ p } - \mathbf{ p }') \delta_{s s'} 
\label {{a,a}}
\eeq
relations.

\subsection{About Fields\label{About Fields}}
A field is
a linear combination of these
annihilation and creation operators 
\beq
\psi_\ell(x) = \sum_s \int \! \frac{d^3p}{(2\pi)^{3/2}} \left[
u_\ell(\mathbf{p},s) e^{ipx} a(\mathbf{ p },s)
+ v_\ell(\mathbf{p},s) e^{-ipx} a^\dagger(\mathbf{ p },s) \right]
\label {psi a a*}
\eeq
that transforms under a Lorentz transformation \( L \) 
followed by a translation by \( a \) as 
\beq
U(L,a) \, \psi_\ell(x) \, U^{-1}(L,a) = 
\sum_{\ell'} D_{\ell \ell'}(L^{-1}) \, \psi_{\ell'}(Lx+a)
\label {Lt rule}
\eeq
and under a rotation \(R\) as
\beq
U(R) \, \psi_\ell(x) \, U^{-1}(R) = 
\sum_{\ell'} D_{\ell \ell'}(R^{-1}) \, \psi_{\ell'}(Rx)
\label {rot rule}
\eeq
in which the operators  \( U(L) \) and \( U(R) \) 
and the matrix \( D(R) \) are unitary~\cite{WeinbergI5}\@.
\par
The mean value
in the vacuum of the product of two components
\( \psi_\ell(x) \) and \( \psi_{\ell'}(y) \) 
of the same field at two space-like points \( x \) and \( y \)
is not identically zero
\beq
\langle 0 | \psi_\ell(x) \, \psi_{\ell'}(y) | 0 \rangle =
\sum_s \int \! \frac{d^3p}{(2\pi)^3} \, 
u_\ell(\mathbf{p},s) 
v_\ell(\mathbf{p},s) e^{ip(x-y)} \ne 0
\label {vac psi psi' vac not 0}
\eeq
as follows explicitly from Eqs.(\ref{[a,a]} \& \ref{{a,a}}) 
for scalar
\beq
\langle 0 | \psi(x) \, \psi(y) | 0 \rangle = 
 \int \! \frac{d^3p}{2p^0 (2\pi)^3} \, e^{ip(x-y)} 
= \Delta_+(x-y) \ne 0
\label {scalar vac psi psi' vac not 0}
\eeq
Dirac
\beq
\langle 0 | \psi_\ell(x) \, \psi_{\ell'}(y) | 0 \rangle = 
i \, [(m - \not \! \partial) 
\gamma^0]_{\ell \ell'} \,
\Delta_+(x-y) \ne 0 .
\label {Dirac vac psi psi' vac not 0}
\eeq
and vector 
\beq
\langle 0 | \psi_\ell(x) \, \psi_{\ell'}(y) | 0 \rangle = 
\left(\delta_{\ell \ell'} 
- \frac{\partial_\ell \partial_{\ell'}}{m^2} \right)
\Delta_+(x-y) \ne 0
\label {vector vac psi psi' vac not 0}
\eeq
fields~\cite{WeinbergI5}\@.  For \( (x - y)^2 = r^2 > 0 \),
the invariant function \( \Delta_+(x - y) \) is~\cite{WeinbergI5}
the modified Bessel function
\beq
\Delta_+(x - y) =  \frac{m}{4\pi^2 r} \,
K_1(m r) > 0 
\label {Delta+}
\eeq
which is positive for \( r > 0 \)\@.

\section{Scalar Fields\label{spin 0}}
Let us consider a scalar field \(\psi(x)\)
at two space-like points \(x_1\) and \(x_2\)\@.
We may choose an inertial coordinate system 
in which these points have the same time
coordinate and lie on the \(x\)-axis
with the spatial origin midway between them.
If the two coordinate systems are related
by the Lorentz transformation \( L \)
followed by the translation by \( a \),
then the points are 
\beq
Lx_1 + a = x_+ = (r/2,0,0,t) 
\;\;\; \mbox{\&} \;\;\>
Lx_2 + a = x_- = (-r/2,0,0,t).
\label {x+x-}
\eeq
\par
For a scalar field \(\psi(x)\), 
the matrix \( D_{\ell \ell'}(L^{-1}) \)
is just the number 1, and so 
the Poincar\'{e} transformation (\ref{Lt rule}) 
reduces to
\beq
U(L,a) \, \psi (x) \, U^{-1}(L,a) =  \psi (Lx+a).
\label {j=0 Lt rule}
\eeq
Using the abbreviation
\( U_{L,a} = U(L,a) \),
inserting three factors
of unity in the form \( I = U_{L,a}^{-1} U_{L,a} \),
and using the invariance of the vacuum state
\( | 0 \rangle \) under Poincar\'{e} transformations
(\( U_{L,a} | 0 \rangle =  | 0 \rangle \)),
we see that the mean value in the vacuum of
\( \psi(x_1) \psi(x_2) \) is the same
as that of \( \psi(x_+) \psi(x_-) \) 
\bea
\langle 0 | \psi(x_1) \psi(x_2)| 0 \rangle & = &
\langle  0 | U_{L,a}^{-1} U_{L,a} \psi(x_1) 
U_{L,a}^{-1} U_{L,a} \psi(x_2) U_{L,a}^{-1} 
U_{L,a} | 0 \rangle \nn\\
& = &
\langle  0 | U_{L,a} \psi(x_1) 
U_{L,a}^{-1} U_{L,a} \psi(x_2) U_{L,a}^{-1} 
| 0 \rangle \nn\\
& = &
\langle 0 | \psi(x_+) \psi(x_-)| 0 \rangle .
\label {j=0 P transf}
\eea
Thus we easily may switch
back and forth between the two coordinate systems.
\par
For a scalar field, the matrix \(  D(R) \),
like the matrix \( D(L) \), is unity, 
and so the general form (\ref{rot rule}) 
of a rotation is simply
\beq
U(R) \, \psi (x) \, U^{-1}(R) =  \psi (Rx).
\label {j=0 rot rule}
\eeq
A rotation by angle \( \pi \) about the \( z \)-axis
takes \( x_+ \) into \( x_- \)
and  \( x_- \) into \( x_+ \)\@.
So such a rotation interchanges
\( \psi(x_+) \) and \( \psi(x_-) \) 
\beq
U_\pi \psi(x_\pm) U^{-1}_\pi = \psi(x_\mp).
\label {pirotj=0}
\eeq
We can use \( U_\pi \) to relate 
\( \langle 0 | \psi(x_+) \, \psi(x_-) | 0 \rangle \)
to \( \langle 0 | \psi(x_-) \, \psi(x_+) | 0 \rangle \)
by thrice inserting unity in the form 
\( I =  U_\pi^{-1}  U_\pi \)
and by using the invariance of 
the vacuum state under rotations 
(\( U_\pi | 0 \rangle =  | 0 \rangle \))
\bea
\langle 0 | \psi(x_+) \, \psi(x_-) | 0 \rangle & = &
\langle 0 |  U_\pi^{-1}  U_\pi \psi(x_+) \,  
U_\pi^{-1}  U_\pi 
\psi(x_-)  U_\pi^{-1}  U_\pi | 0 \rangle \nn\\
& = &
\langle 0 |  U_\pi \psi(x_+) \,  U_\pi^{-1}  U_\pi 
\psi(x_-)  U_\pi^{-1} | 0 \rangle \nn\\
& = &
\langle 0 | \psi(x_-) \, \psi(x_+) | 0 \rangle .
\label {3UdagU}
\eea
Thus the vacuum matrix element
of the commutator \( [ \psi(x_+) , \psi(x_-) ] \)
vanishes as Wigner noted~\cite{Wigner1962} 
\beq
\langle 0 | [ \psi(x_+) , \psi(x_-) ]  | 0 \rangle = 0 
\label {j=0 vev of [] vanishes}
\eeq
but that of the 
anti-commutator \( \{ \psi(x_+) , \psi(x_-) \} \)
is twice that of the product
\beq
\langle 0 | \{ \psi(x_+) , \psi(x_-) \}  | 0 \rangle = 
2 \langle 0 | \psi(x_+) \, \psi(x_-)  | 0 \rangle \ne 0
\label {j=0 vev {} does not vanish}
\eeq
and does not vanish by (\ref{scalar vac psi psi' vac not 0} \&
\ref{Delta+})\@.
So the field \( \psi \) does not
anti-commute with itself at space-like
separations.
\par
What about the creation and annihilation operators?
If they obeyed anti-commutation relations,
the commutator \( [ \psi(x_+) , \psi(x_-) ] \)
would be quadratic in the creation and annihilation operators,
but its mean value in the vacuum still would vanish.
So we can't immediately conclude that they must obey
commutation relations just because 
the mean value (\ref{j=0 vev of [] vanishes})
is zero.
\par
To see whether \( a \) and  \( a^\dagger \) obey commutation
or anti-commutation relations, 
we apply the preceding argument 
(\ref{j=0 Lt rule}--\ref{j=0 vev of [] vanishes}) 
to any product of scalar fields 
at points that can be mapped by a Poincar\'{e}
transformation to \(x_\pm\) and points \( w_k = (0,0,z_k,t_k) \) 
on the \(z\) axis at \textit{arbitrary times}.  
We thus find that mean value in the vacuum
of the commutator
\( [ \psi(x_+) , \psi(x_-) ] \) sandwiched between 
any two such products of fields \( \psi(w_k) \)
\beq
\langle 0 | \psi(w_1) \dots \psi(w_N) \,
[ \psi(x_+) , \psi(x_-) ]  
\, \psi(w_{N+1}) \dots \psi(w_{N+M}) | 0 \rangle = 0 
\label {j=0 vev of fff [] ffff vanishes}
\eeq
must vanish.
If the creation and annihilation operators 
obeyed anti-commutation relations, 
then the commutator \( [ \psi(x_+) , \psi(x_-) ] \)
would be quadratic in them, and 
the Wightman~\cite{PhysRev.101.860} functions 
(\ref{j=0 vev of fff [] ffff vanishes})
would not vanish identically
for arbitrary \( w_k, N \), and \(M\)\@.  
So the creation and annihilation operators can't obey 
anti-commutation relations.
Since by (\ref{[a,a]} \& \ref{{a,a}}) they 
must obey either commutation
or anti-commutation relations, it follows
that they must obey commutation relations.
In this case, the commutator \( [ \psi(x_+) , \psi(x_-) ] \)
is a number, not an operator, and the vanishing
of its vacuum mean value (\ref{j=0 vev of [] vanishes}) 
means that the commutator itself must vanish
\beq
[ \psi(x_+) , \psi(x_-) ] = 0 
\label {j=0 commVan}
\eeq
at the equal-time points \( x_\pm \)\@.
\par
Using the inverse of the 
Poincar\'{e} transformation (\ref{j=0 P transf}),
we find that the commutator 
\( [ \psi(x_1) , \psi(x_2) ] \)
also vanishes for arbitrary space-like
points \( x_1 \) and \( x_2 \)
\bea
[ \psi(x_1) , \psi(x_2) ] & = & 
[ U_{L,a}^{-1} \psi(x_+) U_{L,a},
 U_{L,a}^{-1} \psi(x_-) U_{L,a} ] \nn\\
& = & U_{L,a}^{-1} [ \psi(x_+) , \psi(x_-) ] 
U_{L,a} = U_{L,a}^{-1} 0 U_{L,a} = 0.
\label {j=0 sl veev 0}
\eea
\par
Thus our basic 
assumptions (\ref{phase}--\ref{Delta+})
about creation and annihilation operators
and about how scalar fields transform 
under rotations (and Lorentz transformations)
imply that spin-zero creation and annihilation operators
satisfy commutation relations (\ref{[a,a]}) and that
scalar fields commute at space-like separations (\ref{j=0 sl veev 0})\@.

\section{Majorana and Dirac Fields\label{spin 1/2}}
A Lorentz transformation \( L \)
followed by a translation by \( a \) maps 
a four-component Dirac or Majorana field into
\beq
U(L,a) \, \psi_\ell(x) \, U^{-1}(L,a) = 
\sum_{\ell'=1}^4 D_{\ell \ell'}^{(1/2)}(L^{-1}) \, \psi_{\ell'}(Lx+a)
\label {1/2 Lt rule}
\eeq
in which 
(with suitable \( \gamma \) matrices~\cite{WeinbergI5})
the matrix \(  D(L^{-1})^{(1/2)} \) 
is a \( 4 \times 4 \)
block-diagonal matrix
\beq
D(L^{-1})^{(1/2)} = \pmatrix{ \exp(\vec z \cdot \vec \sigma )& 0 \cr 0 & 
\exp(- \vec z^* \cdot \vec \sigma) }.
\label {1/2 4x4 D(L)}
\eeq
Here the \(\sigma\)-matrices are the Pauli matrices.
Thus under Poincar\'{e} transformations,
the upper two components of the field mix and the lower two
mix, but the upper two components  
do not mix with the lower two.
This holds in particular for  
the transformation \( L, a \)
that takes the space-like points \(x_1\) and \( x_2 \)
to the equal-time points \( x_\pm \)\@.
So to show that the upper or the lower two field components
anti-commute at space-like separations,
we need only show that the upper or the lower two components 
anti-commute at equal times.
We need not show that the upper components anti-commute 
with the lower components at equal times.
\par
Under a rotation \( R \), the rule (\ref{rot rule}) is 
\beq
U(R) \, \psi_\ell(x) \, U^{-1}(R) = 
\sum_{\ell'=1}^4 D_{\ell \ell'}^{(1/2)}(R^{-1}) 
\, \psi_{\ell'}(Rx)
\label {1/2 rot rule}
\eeq
in which \( \vec z \) is imaginary,
and so the two \( 2 \times 2 \) matrices \(r\)
that appear in the \( 4 \times 4 \)
block-diagonal matrix
\beq
D(R^{-1})^{(1/2)} = \pmatrix{ r & 0 \cr 0 & r }
\label {4x4 D(R)}
\eeq
are the same.
The matrix \( r \) that represents 
a right-handed active rotation \(R^{-1}\)
of angle \( \theta = | \vec \theta | \) about the axis \(  \vec \theta \)
is 
\beq
r = \exp( -i \vec \frac{\sigma}{2} \cdot \vec \theta ) 
= \cos (\theta/2) -i \,
\vec \sigma \cdot \hat \theta \, \sin (\theta/2) .
\label {2x2 d(r)}
\eeq
\par
We'll need two special rotations.
The matrices that represent
rotations of \( \pi \)
about the \(z\)- and \(y\)-axes are
\bea
r(\pi,z) & = & \mbox{} -i \sigma_3 = \pmatrix{ -i & 0 \cr 0 & i } 
\label {r(pi,z)} \\
r(\pi,y) & = & \mbox{} -i \sigma_2 = \pmatrix{ 0 & -1 \cr 1 & 0 } .
\label {r(pi,y)} 
\eea
\par
Let \(\psi_\ell\) be a four-component
spin-one-half Majorana or Dirac field.
The transformation rule (\ref {rot rule}) 
tells us that under a left-handed rotation
\(U_{\pi, z}\) of angle \(\pi\)
about the \(z\)-axis, the first and third
components of this field will transform as
\beq
U_{\pi, z} \psi_\ell (x_\pm) U_{\pi, z}^{-1} = 
-i  \psi_\ell (x_\mp) \qquad \ell = 1, 3
\label {j=1/2 13z}
\eeq
since the rotation \(R^{-1}\) is a 
right-handed active rotation about the \(z\)-axis
by angle \(\pi\)\@.
The second and fourth components will transform as
\beq
U_{\pi, z} \psi_\ell (x_\pm) U_{\pi, z}^{-1} = 
i  \psi_\ell (x_\mp)  \qquad \ell = 2, 4.
\label {j=1/2 24z}
\eeq
\par
Similarly, under a left-handed rotation
\(U_{\pi, y}\) of angle \(\pi\)
about the \(y\)-axis, the first and third
components of this field will transform as
\beq
U_{\pi, y} \psi_\ell (x_\pm) U_{\pi, y}^{-1} = 
\psi_{\ell+1} (x_\mp)  \qquad \ell = 1, 3
\label {j=1/2 13y}
\eeq
since the rotation \(R^{-1}\) is a 
right-handed active rotation about the \(y\)-axis
by angle \(\pi\)\@.
The second and fourth components will transform as
\beq
U_{\pi, y} \psi_\ell (x_\pm) U_{\pi, y}^{-1} = 
-  \psi_{\ell-1} (x_\mp) \qquad \ell = 2, 4.
\label {j=1/2 24y}
\eeq
\par
Let us first consider the case in which 
the two indices \(\ell\) and \(\ell'\) are the same.
The vacuum state is invariant under rotations,
so if we thrice insert unity in the form 
\( I =  U_{\pi,z}^{-1}  U_{\pi, z} \), 
and use Eqs.(\ref {j=1/2 13z} \& \ref{j=1/2 24z}),
then (not summing over \(\ell\)) we get
\bea
\langle 0 | \psi_\ell (x_+) \, \psi_\ell (x_-) | 0 \rangle & = &
\langle 0 |  U_{\pi, z}^{-1}  U_{\pi, z} \psi_\ell(x_+) \,  
U_{\pi, z}^{-1}  U_{\pi, z} 
\psi_\ell(x_-)  U_{\pi, z}^{-1}  U_{\pi, z} | 0 \rangle \nn\\
& = &
\langle 0 |  U_{\pi, z} \psi_\ell(x_+) \,  U_{\pi, z}^{-1}  U_{\pi, z} 
\psi_\ell(x_-)  U_{\pi, z}^{-1} | 0 \rangle \nn\\
& = &
- \langle 0 | \psi_\ell(x_-) \, \psi_\ell(x_+) | 0 \rangle .
\label {j=1/2 3UdagU}
\eea
Thus the vacuum matrix element
of the anti-commutator \( \{ \psi(x_+) , \psi(x_-) \} \)
vanishes
\beq
\langle 0 | \{ \psi_\ell(x_+) , \psi_\ell(x_-) \}  | 0 \rangle = 0.
\label {j=1/2 vev of [] vanishes}
\eeq
But that of the 
commutator \( [ \psi_\ell(x_+) , \psi_\ell(x_-) ]\)
is twice that of the product
\beq
\langle 0 | [ \psi_\ell(x_+) , \psi_\ell(x_-) ]  | 0 \rangle = 
2 \langle 0 | \psi_\ell(x_+) \, \psi_\ell(x_-)  | 0 \rangle 
\label {j=1/2 vev {} does not vanish}
\eeq
which by (\ref{Dirac vac psi psi' vac not 0}
\& \ref{Delta+}) does not vanish identically.
So the vacuum matrix element (\ref{j=1/2 vev {} does not vanish})
of the commutator \( [ \psi_\ell(x_+) , \psi_\ell(x_-) ]\)
does not vanish;
the fields \(\psi_\ell(x_+)\) and \( \psi_\ell(x_-) \) 
do not commute.  
\par
What about unequal indices?
If we thrice insert 
unity in the form 
\( I =  U_{\pi,y}^{-1}  U_{\pi, y} \), 
and use the invariance of the vacuum under rotations
as well as Eqs.(\ref {j=1/2 13y} \& \ref{j=1/2 24y}),
then we find
\bea
\langle 0 | \psi_1 (x_+) \, \psi_2 (x_-) | 0 \rangle & = &
\langle 0 |  U_{\pi, y} \psi_1(x_+) \,  U_{\pi, y}^{-1}  U_{\pi, y} 
\psi_2(x_-)  U_{\pi, y}^{-1} | 0 \rangle \nn\\
& = &
- \langle 0 | \psi_2(x_-) \, \psi_1(x_+) | 0 \rangle .
\label {j=1/2 12 3UdagU}
\eea
So the mean value in the vacuum 
of the anti-commutator 
\beq
\langle 0 | \{ \psi_1(x_+) , \psi_2(x_-) \} | 0 \rangle = 0 .
\label {j=1/2 12 anticommVan}
\eeq
vanishes, but by (\ref{Dirac vac psi psi' vac not 0}
\& \ref{Delta+})
not that of the commutator 
\beq
\langle 0 | [ \psi_1(x_+) , \psi_2(x_-) ] | 0 \rangle \ne 0 .
\label {j=1/2 12 commVan}
\eeq
So the fields \(\psi_1(x_+)\) and \( \psi_2(x_-) \) 
do not commute. 
\par
An identical argument shows that
the vacuum mean value of the anti-commutator 
\beq
\langle 0 | \{ \psi_3(x_+) , \psi_4(x_-) \} | 0 \rangle = 0 .
\label {j=1/2 34 anticommVan}
\eeq
vanishes, but not that of the commutator 
\beq
\langle 0 | [ \psi_3(x_+) , \psi_4(x_-) ] | 0 \rangle \ne 0 .
\label {j=1/2 34 commVan}
\eeq
So the fields \(\psi_3(x_+)\) and \( \psi_4(x_-) \) 
do not commute. 
\par
What about the creation and annihilation operators?
To see whether they obey commutation
or anti-commutation relations, 
we apply the rotation \( U_{\pi,z} \) 
(\ref{r(pi,z)}, \ref{j=1/2 13z}) and 
the preceding argument (\ref{j=1/2 24z},
\ref{j=1/2 3UdagU}, \& \ref{j=1/2 vev of [] vanishes}) 
to any product of Majorana or Dirac fields 
at points that can be mapped by a Poincar\'{e}
transformation to \(x_\pm\) and points \( w_k = (0,0,z_k,t_k) \) 
on the \(z\) axis at \textit{arbitrary times}.
We thus find that the anti-commutator
\( \{ \psi_\ell(x_+) , \psi_{\ell}(x_-) \} \) 
of fields of the same index
sandwiched between 
any two such products of fields \( \psi_{\ell_k}(w_k) \)
\beq
\langle 0 | \left( \prod_{k=1}^N \psi_{\ell_k}(w_k) \right)
\{ \psi_{\ell}(x_+) , \psi_{\ell}(x_-) \} 
\left( \prod_{k=N+1}^{N+M} \psi_{\ell_k}(w_k) \right)
| 0 \rangle = 0 
\label {j=1/2 vev of fff [] ffff vanishes}
\eeq
vanishes as long as the numbers \( N_{\ell_k} + M_{\ell_k} \) 
of fields of index \(\ell_k\) 
(not including 
\(  \psi_{\ell}(x_+) \) \& \( \psi_{\ell'}(x_-) \) )
satisfy
\beq 
N_1 +M_1 + N_3 + M_3 = N_2 + M_2 + N_4 + M_4 \pm 4 n 
\label {4n}
\eeq
for some integer \(n\)\@.
If the creation and annihilation operators 
obeyed commutation relations, 
then the anti-commutator \( \{ \psi(x_+) , \psi(x_-) \} \)
would be quadratic in them, and so 
the Wightman~\cite{PhysRev.101.860} 
functions (\ref{j=1/2 vev of fff [] ffff vanishes})
could not vanish
for arbitrary \( w_k, N \), and \(M\) satisfying (\ref{4n})\@.  
So they can't obey 
commutation relations.
Since by (\ref{[a,a]} \& \ref{{a,a}}) they 
must obey either commutation
or anti-commutation relations, it follows
that they must obey anti-commutation relations.
In this case, the anti-commutator \( \{ \psi(x_+) , \psi(x_-) \} \)
is a number, not an operator, and the vanishing
of its vacuum mean value (\ref{j=1/2 vev of [] vanishes}) 
means that the anti-commutator itself must vanish
\beq
\{ \psi_\ell(x_+) , \psi_\ell(x_-) \} = 0 .
\label {j=1/2 commVan}
\eeq
Finally, by using the relation
(\ref{j=0 sl veev 0})
(adapted via (\ref{1/2 4x4 D(L)}) to Dirac fields),
we conclude that 
at space-like separations,
the upper two field components anti-commute,
and so do the lower two.
\par
What about the anti-commutator 
\( [ \psi_1(x_+) , \psi_3(x_-) ]_+  \)?
Parity relates fields with indices 1 and 2 
to those with indices 3 and 4\@.
Under parity, a free Majorana field 
transforms as
\beq
P \psi(x) P^{-1} = \eta^* \beta \psi(\mathcal{P}x)
\label {parity}
\eeq
in which \( \beta = i \gamma^0 \)\@. 
Its intrinsic parity 
\beq
\eta = \pm i
\label {eta}
\eeq
is imaginary~\cite{WeinbergI5}\@.
So for \( \ell = 1 \) or 2,
\beq
P \psi_\ell(x_\pm) P^{-1} = \eta^* \psi_{\ell+2}(x_\mp)
\label {parityPsi1,2}
\eeq
and for  \( \ell = 3 \) or 4,
\beq
P \psi_\ell(x_\pm) P^{-1} = \eta^* \psi_{\ell-2}(x_\mp)
\label {parityPsi3,4}
\eeq
Thus if we thrice insert 
unity in the form \( P P^{-1} \) 
and use the invariance under parity of the vacuum
of the free field theory,
then we get 
\bea
\langle 0 | \psi_1 (x_+) \, \psi_3 (x_-) | 0 \rangle & = &
\langle 0 | P \psi_1(x_+) P^{-1}  P 
\psi_3(x_-) P^{-1} | 0 \rangle \nn\\
& = &
(\eta^*)^2 \langle 0 | \psi_3(x_-) \, \psi_1(x_+) | 0 \rangle \nn\\
& = &
- \langle 0 | \psi_3(x_-) \, \psi_1(x_+) | 0 \rangle
\label {j=1/2 13 PP}
\eea
since \(\eta\)
is imaginary.
So the mean value in the vacuum
of the anti-commutator 
\beq
\langle 0 | \{ \psi_1(x_+) , \psi_3(x_-) \} | 0 \rangle = 0 
\label {j=1/2 13 commVan}
\eeq
vanishes.  And since the
creation and annihilation operators 
obey anti-commutation relations,
the anti-commutator 
\( \{ \psi_1(x_+) , \psi_3(x_-) \} \) 
is a number, not an operator.
So the vanishing of its mean value
in the vacuum implies that the anti-commutator 
itself vanishes
\beq
\{ \psi_1(x_+) , \psi_3(x_-) \} = 0 .
\label {j=1/2 13 {}Van}
\eeq
An identical argument shows that
the anti-commutator vanishes
\beq
\{ \psi_2(x_+) , \psi_4(x_-) \} = 0 .
\label {j=1/2 24 {}Van}
\eeq
Since Dirac fields are complex linear combinations
of two Majorana fields of the same mass,
Eqs(\ref{j=1/2 13 {}Van} \& \ref{j=1/2 24 {}Van})
apply also to Dirac fields.
\par
Thus our basic assumptions 
(\ref{phase}--\ref{Delta+})
about creation and annihilation operators
and about how Majorana and Dirac fields transform
under rotations (and Lorentz transformations
and parity)
imply that spin-one-half creation and annihilation operators
satisfy anti-commutation relations (\ref{{a,a}}) and that
Majorana and Dirac fields anti-commute at space-like separations.

\section{Vector Fields\label{spin 1}}
A vector field transforms
like a 3-vector under rotations
\beq
U(R) \, \psi_\ell(x) \, U^{-1}(R) = 
\sum_{\ell'=1}^3 D_{\ell \ell'}^{(1)}(R^{-1}) \, \psi_{\ell'}(Rx) 
= \sum_{\ell'=1}^3 R^{-1}_{\ell \ell'} \, \psi_{\ell'}(Rx) 
\label {1 rot rule}
\eeq
since
the matrix \( D(R^{-1})^{(1)} \) that appears
in the transformation rule (\ref{rot rule})
is simply the \( 3 \times 3 \) matrix \( R^{-1} \)\@.
The matrix of a  right-handed rotation \( R^{-1} \) by angle \( \theta \)
about the axis \( \hat n \) is~\cite{Cahill2000}
\beq
R_{ij}^{-1}(\theta \hat n) = \delta_{ij} \, \cos \theta 
- \sin \theta \sum_{k = 1}^3 \epsilon_{ijk} \hat n_k 
+ ( 1 - \cos \theta) \hat n_i \hat n_j
\label {3x3 R formula}
\eeq
in which \( \epsilon_{ijk} \) is totally anti-symmetric
in \( i, j, k \)\@.
\par
To deal with the various pairs of 
indices \( \ell \) and \( \ell' \), 
we rotate the reference points
from \(x_\pm = ( \pm x, 0, 0, t ) \)
to \( y_\pm = ( \pm y, \pm y, \pm y, t ) \)
where \( y = x/\sqrt{3} \)\@.
The space parts of the 4-vectors \( y_\pm \) 
are parallel or anti-parallel 
to the 3-vector \( \mathbf{r} = ( 1, 1, 1) \)\@.
Any rotation by \( \pi \) about
any axis \( \hat n \) that is perpendicular
to the vector \( \mathbf{r} \)
will interchange  \( y_\pm \)
with \( y_\mp \)\@.  
Any such axis satisfies \( \hat n \cdot \mathbf{r} = 0 \)
and so is of the form
\beq
\hat n = ( a - b, b - c, c - a)
\label {hat n}
\eeq
with \( ( a - b)^2 + (b - c)^2 +(c - a)^2 = 1 \)\@.
A rotation  \( R(\pi \hat n) =  R^{-1}(\pi \hat n) \) 
by \(\pi\) about the axis \( \hat n \) 
is represented by the \( 3 \times 3 \) matrix
\beq
\! \! \!
R^{-1}(\pi \hat n) = \pmatrix{ 2(a-b)^2 - 1 & 2(a-b)(b-c) & 2(a-b)(c-a) \cr
2(b-c)(a-b) & \! 2(b-c)^2 - 1 & 2(b-c)(c-a) \cr
2(c-a)(a-b) & 2(c-a)(b-c) & 2(c-a)^2 - 1 }
\label {R(pi hat n)}
\eeq
according to the general formula (\ref{3x3 R formula})\@.
\par
We shall use three special cases.
A rotation by \( \pi \)
about the axis \( \hat n_1 = ( 1/\sqrt{2}, - 1/\sqrt{2}, 0 ) \) is 
represented by the matrix
\beq
R^{-1}(\pi \hat n_1) = \pmatrix{ 0 & -1 & 0 \cr
                                -1 & 0 & 0 \cr
                                 0 & 0 & -1 }.
\label {R n1}
\eeq
By the rule (\ref{rot rule}), the unitary operator
\( U_{\pi,\hat n_1} \) 
rotates the fields \( \psi_1(\pm y) \)
and \( \psi_2(\pm y) \) into \( - \psi_2(\mp y) \) 
and  \( - \psi_1(\mp y) \), and the field 
\( \psi_3(\pm y) \) into \( - \psi_3(\mp y) \) 
\bea
U_{\pi,\hat n_1} \psi_1(y_\pm) U_{\pi,\hat n_1}^{-1} & = &
R^{-1}_{1 \ell'} (\pi \hat n_1)  \psi_{\ell'}(y_\mp) =
- \psi_2(y_\mp) \nn\\
U_{\pi,\hat n_1} \psi_2(y_\pm) U_{\pi,\hat n_1}^{-1} & = &
R^{-1}_{2 \ell'} (\pi \hat n_1)  \psi_{\ell'}(y_\mp) =
- \psi_1(y_\mp) \nn\\
U_{\pi,\hat n_1} \psi_3(y_\pm) U_{\pi,\hat n_1}^{-1} & = &
R^{-1}_{3 \ell'} (\pi \hat n_1)  \psi_{\ell'}(y_\mp) =
- \psi_3(y_\mp) .
\label {R n1 123}
\eea
\par
A rotation by \( \pi \)
about the axis \( \hat n_2 = ( 0 , 1/\sqrt{2}, - 1/\sqrt{2} ) \) is 
represented by the matrix
\beq
R^{-1}(\pi \hat n_2) = \pmatrix{ -1 & 0 & 0 \cr
                                 0 & 0 & -1 \cr
                                 0 & -1 & 0 }.
\label {R n2}
\eeq
By the rule (\ref{rot rule}), the unitary operator
\( U_{\pi,\hat n_2} \) 
rotates the fields \( \psi_2(\pm y) \)
and \( \psi_3(\pm y) \) into \( - \psi_3(\mp y) \) 
and  \( - \psi_2(\mp y) \), and the field 
\( \psi_1(\pm y) \) into \( - \psi_1(\mp y) \) 
\bea
U_{\pi,\hat n_2} \psi_1(y_\pm) U_{\pi,\hat n_2}^{-1} & = &
R^{-1}_{1 \ell'} (\pi \hat n_2)  \psi_{\ell'}(y_\mp) =
- \psi_1(y_\mp) \nn\\
U_{\pi,\hat n_2} \psi_2(y_\pm) U_{\pi,\hat n_2}^{-1} & = &
R^{-1}_{2 \ell'} (\pi \hat n_2)  \psi_{\ell'}(y_\mp) =
- \psi_3(y_\mp) \nn\\
U_{\pi,\hat n_2} \psi_3(y_\pm) U_{\pi,\hat n_2}^{-1} & = &
R^{-1}_{3 \ell'} (\pi \hat n_2)  \psi_{\ell'}(y_\mp) =
- \psi_2(y_\mp) .
\label {R n2 123}
\eea
\par
A rotation by \( \pi \)
about the axis \( \hat n_3 = ( 1/\sqrt{2}, 0, - 1/\sqrt{2} ) \) is 
represented by the matrix
\beq
R^{-1}(\pi \hat n_3) = \pmatrix{ 0 & 0 & -1 \cr
                                 0 & -1 & 0 \cr
                                -1 & 0 & 0 }.
\label {R n3}
\eeq
By the rule (\ref{rot rule}), the unitary operator
\( U_{\pi,\hat n_3} \) 
rotates the fields \( \psi_1(\pm y) \),
\( \psi_2(\pm y) \), and \( \psi_3(\pm y) \)  
into \( - \psi_3(\mp y) \) 
and  \( - \psi_2(\mp y) \), and  
\( - \psi_1(\mp y) \) 
\bea
U_{\pi,\hat n_3} \psi_1(y_\pm) U_{\pi,\hat n_3}^{-1} & = &
R^{-1}_{1 \ell'} (\pi \hat n_3)  \psi_{\ell'}(y_\mp) =
- \psi_3(y_\mp) \nn\\
U_{\pi,\hat n_3} \psi_2(y_\pm) U_{\pi,\hat n_3}^{-1} & = &
R^{-1}_{2 \ell'} (\pi \hat n_3)  \psi_{\ell'}(y_\mp) =
- \psi_2(y_\mp) \nn\\
U_{\pi,\hat n_3} \psi_3(y_\pm) U_{\pi,\hat n_3}^{-1} & = &
R^{-1}_{3 \ell'} (\pi \hat n_3)  \psi_{\ell'}(y_\mp) =
- \psi_1(y_\mp) .
\label {R n3 123}
\eea
\par
The transformations (\ref{R n1 123}--\ref{R n3 123})
and the invariance of the vacuum under rotations
imply these relations:
\bea
\langle 0 | \psi_1 (y_\pm) \, \psi_1 (y_\mp) | 0 \rangle & = &
\langle 0 | U_{\pi, \hat n_2} \psi_1(y_\pm) \,  
U_{\pi, \hat n_2}^{-1}  U_{\pi, \hat n_2} 
\psi_1(y_\mp)  U_{\pi, \hat n_2}^{-1} | 0 \rangle \nn\\
& = & \langle 0 | \psi_1 (y_\mp) \, \psi_1 (y_\pm) | 0 \rangle \nn\\
\langle 0 | \psi_1 (y_\pm) \, \psi_2 (y_\mp) | 0 \rangle & = &
\langle 0 | U_{\pi, \hat n_1} \psi_1(y_\pm) \,  
U_{\pi, \hat n_1}^{-1}  U_{\pi, \hat n_1} 
\psi_2(y_\mp)  U_{\pi, \hat n_1}^{-1} | 0 \rangle \nn\\
& = & \langle 0 | \psi_2 (y_\mp) \, \psi_1 (y_\pm) | 0 \rangle \nn\\
\langle 0 | \psi_1 (y_\pm) \, \psi_3 (y_\mp) | 0 \rangle & = &
\langle 0 | U_{\pi, \hat n_3} \psi_1(y_\pm) \,  
U_{\pi, \hat n_3}^{-1}  U_{\pi, \hat n_3} 
\psi_3(y_\mp)  U_{\pi, \hat n_3}^{-1} | 0 \rangle \nn\\
& = & \langle 0 | \psi_3 (y_\mp) \, \psi_1 (y_\pm) | 0 \rangle \nn\\
\langle 0 | \psi_2 (y_\pm) \, \psi_2 (y_\mp) | 0 \rangle & = &
\langle 0 | U_{\pi, \hat n_3} \psi_2(y_\pm) \,  
U_{\pi, \hat n_3}^{-1}  U_{\pi, \hat n_3} 
\psi_2(y_\mp)  U_{\pi, \hat n_3}^{-1} | 0 \rangle \nn\\
& = & \langle 0 | \psi_2 (y_\mp) \, \psi_2 (y_\pm) | 0 \rangle \nn\\
\langle 0 | \psi_2 (y_\pm) \, \psi_3 (y_\mp) | 0 \rangle & = &
\langle 0 | U_{\pi, \hat n_2} \psi_2(y_\pm) \,  
U_{\pi, \hat n_2}^{-1}  U_{\pi, \hat n_2} 
\psi_3(y_\mp)  U_{\pi, \hat n_2}^{-1} | 0 \rangle \nn\\
& = & \langle 0 | \psi_3 (y_\mp) \, \psi_2 (y_\pm) | 0 \rangle \nn\\
\langle 0 | \psi_3 (y_\pm) \, \psi_3 (y_\mp) | 0 \rangle & = &
\langle 0 | U_{\pi, \hat n_1} \psi_3(y_\pm) \,  
U_{\pi, \hat n_1}^{-1}  U_{\pi, \hat n_1} 
\psi_3(y_\mp)  U_{\pi, \hat n_1}^{-1} | 0 \rangle \nn\\
& = & \langle 0 | \psi_3 (y_\mp) \, \psi_3 (y_\pm) | 0 \rangle .
\label {1 vevs}
\eea
Thus the mean value in the vacuum
of the commutator vanishes
\beq
\langle 0 | [ \psi_i (y_\pm) , \psi_j (y_\mp) ] | 0 \rangle
= 0
\label {1 ij <[,]>=0}
\eeq
for all pairs, \(i,j = 1, 2, 3\),
but by (\ref{vector vac psi psi' vac not 0}
\& \ref{Delta+}) 
that of the anti-commutator does not.
So the fields \( \psi_i (y_\pm) \) and \( \psi_j (y_\mp) \)
do not anti-commute.
\par
To see whether the creation and annihilation operators
for vector fields obey commutation
or anti-commutation relations,
we use a rotation \( U_{\pi,z} \) by angle \( \pi \)
about the \( z \) axis with rotation matrix
\beq
R^{-1}(\pi \hat z) = \pmatrix{ -1 & 0 & 0 \cr
                                 0 & -1 & 0 \cr
                                 0 & 0 & 1 }.
\label {R z}
\eeq
We apply this rotation and 
the preceding argument 
(\ref{1 rot rule}--\ref{1 ij <[,]>=0})
to the fields \( \psi_i(x_+) \psi_i(x_-) \)
sandwiched between products of fields 
at points that can be mapped by a Poincar\'{e}
transformation to \(x_\pm\) and points \( w_k = (0,0,z_k,t_k) \) 
on the \(z\) axis at \textit{arbitrary times}\@.
Thus we find that the 
commutator \( [ \psi_i(x_+), \psi_i(x_-) ] \)
of vector fields of the same index 
sandwiched between 
any two such products of fields \( \psi_{j_k}(w_k) \)
\beq
\langle 0 | \left( \prod_{k=1}^N \psi_{j_k}(w_k) \right)
[ \psi_{i}(x_+) , \psi_{i}(x_-) ] 
\left( \prod_{k=N+1}^{N+M} \psi_{j_k}(w_k) \right)
| 0 \rangle = 0 
\label {j=1 vev of fff [] ffff vanishes}
\eeq
vanishes as long as the numbers of fields
of index 1 or 2 is even
\beq
N_1 + M_1 + N_2 + M_2 = \pm 2n .
\label {1 or 2 even}
\eeq
If the creation and annihilation operators 
obeyed anti-commutation relations, then
the commutator \( [ \psi_i(x_+), \psi_i(x_-) ] \) 
would be quadratic in them, and so 
the Wightman~\cite{PhysRev.101.860} functions  
(\ref{j=1 vev of fff [] ffff vanishes})
could not vanish
for all \( N_k + M_k \) that satisfy (\ref{1 or 2 even})
and for arbitrary \( w_k \)\@.
So they can't obey anti-commutation relations.
Since by (\ref{[a,a]} \& \ref{{a,a}}), they must obey
either commutation
or anti-commutation relations,
they must obey commutation relations.
In this case, the equal-time commutators
\( [ \psi_i (y_\pm) , \psi_j (y_\mp) ] \)
are numbers, not operators, and so the vanishing
of their mean values in the vacuum implies
that these commutators vanish
\beq
[ \psi_i (y_\pm) , \psi_j (y_\mp) ] = 0 .
\label {1 ij [,]}
\eeq
To extend this relation to 
points at arbitrary space-like
separations, we need to know
how the vector field transforms under
Lorentz boosts.
\par
The response of a vector field
to a Lorentz boost depends upon
the mass of the particle the field
describes.  If the particle is massive,
then under a Lorentz transformation \( L \) 
followed by a translation by \( a \) 
the field transforms as 
\beq
U(L,a) \, \psi_\mu(x) \, U^{-1}(L,a) = 
L^\nu_{\ \mu} \, \psi_\nu(Lx+a) .
\label {j=1 m>0 Lt rule}
\eeq
In this case, there is no difficulty
applying the argument (\ref{j=0 sl veev 0}) 
to the vector field \( \psi_\mu \)
and so extending the vanishing
of the equal-time commutator (\ref{1 ij [,]})
to points at arbitrary space-like separations.
\par
But if the field \( \psi_\mu \) 
represents a massless particle,
then under a Lorentz boost the
preceding equation is augmented by
a gauge transformation 
\beq
U(L,a) \, \psi_\mu(x) \, U^{-1}(L,a) = 
L^\nu_{\ \mu} \, \psi_\nu(Lx+a) 
+ \partial_\mu \Omega(x+a,L)
\label {j=1 m=0 Lt rule}
\eeq
in which \( \Omega \) is a linear
combination of annihilation and
creation operators~\cite{WeinbergI5}, and 
I do not know of a simple way of extending
Eq.(\ref{1 ij [,]}) to space-like separations.

\section{More General Fields\label{any spin}}
A field that transforms according to 
the \( (A,B) \) representation of the
homogeneous Lorentz group transforms 
under rotations as~\cite{WeinbergI5}
\beq
U(R) \, \psi_{ab}(x) \, U^{-1}(R) = 
\sum_{a', b'}
D_{a a'}^{(A)}(R^{-1}) \, D_{b b'}^{(B)}(R^{-1}) \,
\psi_{a'b'}(Rx)
\label {(A,B) rot rule}
\eeq
in which the sums run in integer steps
from \( -A \) to \( A \)
and from \( -B \) to \( B \)\@.
This field can represent a particle
of spin \( j \) where 
\( |A-B| \le j \le A + B \) and \( j \)
differs from \( A + B \) by an integer~\cite{WeinbergI5}\@.
If the rotation \( R \) is a right-handed 
rotation of angle \( \theta \) about
the \(z\)-axis, then
\beq
D_{a a'}^{(A)}(R^{-1}) = \delta_{a a'} \, e^{i a \theta}
\quad \mbox{and} \quad
D_{b b'}^{(B)}(R^{-1}) = \delta_{b b'} \, e^{i b \theta}.
\label {j z rot}
\eeq
Thus under a right-handed rotation by \( \pi \)
about the \(z\)-axis, the field \( \psi_{ab}(x_\pm) \)
at \( x_\pm = (\pm x,0,0,t) \) transforms into
\beq
U_{\pi,z} \, \psi_{ab}(x_\pm) \, U^{-1}_{\pi,z} = 
e^{i (a+b) \pi} \,
\psi_{ab}(x_\mp).
\label {(A,B) pi,z rot rule}
\eeq
Again inserting unity in the form \( I = U_{\pi,z} U^{-1}_{\pi,z} \)
and invoking the invariance of the vacuum under rotations,
we find 
\bea
\langle 0 | \psi_{ab} (x_\pm) \, \psi_{ab} (x_\mp) | 0 \rangle & = &
\langle 0 | U_{\pi, z} \psi_{ab}(x_\pm) \,  
U_{\pi, z}^{-1}  U_{\pi, z} 
\psi_{ab}(x_\mp)  U_{\pi, z}^{-1} | 0 \rangle \nn\\
& = & e^{i 2 (a+b) \pi} \langle 0 | \psi_{ab} (x_\mp) \, 
\psi_{ab} (x_\pm) | 0 \rangle .
\eea
Now the indices \( a \) and \( b \) 
are related to \( A \) and \( B \) 
by \( a = A - n \) and \( b = B - m \)
in which \( n \) and \( m \) are both
integers.  Thus
\( 2(a + b) = 2A + 2B - 2n - 2m \),
that is, \( 2(a + b) \) differs from \( 2A + 2B \)
by an even integer, and so,
\( 2(a + b) \) differs from \( 2 j \)
by an even integer.  Thus
\bea
\langle 0 | \psi_{ab} (x_\pm) \, \psi_{ab} (x_\mp) | 0 \rangle 
& = &e^{i 2 (a+b) \pi} \langle 0 | \psi_{ab} (x_\mp) \, 
\psi_{ab} (x_\pm) | 0 \rangle \nn\\
& = & 
e^{i 2 j \pi} \langle 0 | \psi_{ab} (x_\mp) \, 
\psi_{ab} (x_\pm) | 0 \rangle \nn\\
& = &
(-1)^{2j}  \langle 0 | \psi_{ab} (x_\mp) \, 
\psi_{ab} (x_\pm) | 0 \rangle.
\label {ab (-1)^2j}
\eea
Thus for particles of integral spin, bosons,
the mean value in the vacuum of
the commutator vanishes
\beq
\langle 0 | [ \psi_{ab} (x_\pm), \psi_{ab} (x_\mp) ] | 0 \rangle 
= 0
\label {j vac bosons}
\eeq
while that of the anti-commutator does not vanish
\beq
\langle 0 | \{ \psi_{ab} (x_\pm), \psi_{ab} (x_\mp) \} | 0 \rangle 
= 2 \langle 0 | \psi_{ab} (x_\pm) \, \psi_{ab} (x_\mp) | 0 \rangle 
\ne 0 .
\label {j anti vac bosons}
\eeq
Also, for particles of half-odd-integral spin,
fermions, 
the mean value in the vacuum of
the anti-commutator vanishes
\beq
\langle 0 | \{ \psi_{ab} (x_\pm), \psi_{ab} (x_\mp) \} | 0 \rangle 
= 0
\label {j vac fermions}
\eeq
while that of the commutator does not vanish
\beq
\langle 0 | [ \psi_{ab} (x_\pm), \psi_{ab} (x_\mp) ] | 0 \rangle 
= 2 \langle 0 | \psi_{ab} (x_\pm) \, \psi_{ab} (x_\mp) | 0 \rangle 
\ne 0 .
\label {j anti vac fermions}
\eeq
\par
So the fields do not obey the wrong statistics.
To see whether the creation and annihilation operators
for these general fields obey commutation
or anti-commutation relations,
we apply the rotation \( U_{\pi,z} \) 
to the fields \( \psi_{ab}(x_+) \psi_{ab}(x_-) \)
sandwiched between products of fields 
at points \( w_k = (0,0,z_k,t_k) \) 
on the \(z\) axis at \textit{arbitrary times}\@.
Thus we find that the \(j\)-commutator 
\beq
[ \psi_{ab}(x_+),  \psi_{ab}(x_-) ]_j 
= \psi_{ab}(x_+) \psi_{ab}(x_-) - (-1)^{2j} 
\psi_{ab}(x_-) \psi_{ab}(x_+) 
\label {j comm}
\eeq
of two
general fields of the same index 
sandwiched between 
any two such products of fields \( \psi_{a_k b_k}(w_k) \)
\beq
\!\!\!\!\!\!\!\!\!\!
\langle 0 | \left( \prod_{k=1}^N \psi_{a_k b_k}(w_k) \right)
[ \psi_{ab}(x_+) , \psi_{ab}(x_-) ]_j 
\left( \prod_{k=N+1}^{N+M} \psi_{a_k b_k}(w_k) \right)
| 0 \rangle = 0 
\label {ab vev of fff [] ffff vanishes}
\eeq
vanishes as long as 
\beq
\sum_{k=1}^{N+M} a_k + b_k = 2n .
\label {gen ab even}
\eeq
If the creation and annihilation operators 
obeyed the wrong commutation relations, then
the \(j\)-commutator \( [ \psi_i(x_+), \psi_i(x_-) ]_j \) 
would be quadratic in them, and so 
these Wightman~\cite{PhysRev.101.860} functions  
(\ref{ab vev of fff [] ffff vanishes})
could not vanish
for all \( N, M \) and all \(a_k and b_k \)
that satisfy (\ref{gen ab even})
and for arbitrary \( w_k \)\@.
So they can't obey the wrong commutation relations.
Since by (\ref{[a,a]} \& \ref{{a,a}}), they must obey
either commutation
or anti-commutation relations,
they must obey the right commutation relations.
In this case, the equal-time \(j\)-commutators
\( [ \psi_i(x_+), \psi_i(x_-) ]_j \)
are numbers, not operators, and so the vanishing
of their mean values in the vacuum implies
that these commutators vanish
\beq
 [ \psi_{ab}(x_+), \psi_{ab}(x_-) ]_j = 0
\label {ab [,]j}
\eeq
\par
It should be possible to extend this argument
to arbitrary space-like separations
and to all pairs of indices \( ab, a'b' \)\@.

\ack
Thanks to James Lowe, Randolph Reeder, and
Steven Weinberg  
for helpful conversations.
\section*{References}
\bibliography{physics,proteins}
\end{document}